\documentclass[12pt]{amsart}
\usepackage{amssymb,latexsym}
\newtheorem{theorem}{Theorem}

\newtheorem{corollary}{Corollary}

\newtheorem{lemma}{Lemma}

\newcommand{\Rbb}{\mathbb{R}}
\newcommand{\zbar}{\overline{z}}
\newcommand{\npl}{n_{+}}
\newcommand{\nmi}{n_{-}}

\begin{document}

\title{Laguerre-Gaussian Modes and the Wigner Transform}
\author{Michael VanValkenburgh}
\address{UCLA Department of Mathematics, Los Angeles, CA 90095-1555, USA}
\email{mvanvalk@ucla.edu}
\maketitle

\begin{abstract}
Recent developments in laser physics have called renewed attention
to Laguerre-Gaussian (LG) beams of paraxial light. In this paper
we consider the corresponding LG modes for the two-dimensional
harmonic oscillator, which appear in the transversal plane at the
laser beam's waist. We see how they arise as Wigner transforms of
Hermite-Gaussian modes, and we proceed to find a closed form for
their own Wigner transforms, providing an alternative to the
methods of Simon and Agarwal. Our main observation is that the
Wigner transform intertwines the creation and annihilation
operators for the two classes of modes.
\end{abstract}

\vspace{12pt}


\vspace{12pt}

\section{Introduction}

In the early 1990s it was observed that a Laguerre-Gaussian (LG)
beam of paraxial light has a well-defined orbital angular momentum
and that such a beam may be created from a Hermite-Gaussian (HG)
beam by way of an astigmatic optical system \cite{R:ABSW}. This
discovery was soon to find applications, for example, in biology
(for ``optical tweezers'') and in the study of quantum
entanglement \cite{R:OpticalAM}. Additionally, recent experiments
suggest that beams with orbital angular momentum might be
generated \emph{in situ}, in free electron lasers \cite{R:PAC07}.

In this paper we consider the corresponding LG modes for the
two-dimensional harmonic oscillator, which appear in the
transversal plane at the laser beam's waist. Indeed, one may use
the operator algebra of the harmonic oscillator to study the
analytical forms of the HG and LG beams \cite{R:NA}, \cite{R:VN}.
Here we demonstrate that the LG modes arise as Wigner transforms
of HG modes, and we proceed to find their own Wigner transforms.
Our methods provide an alternative to those of Gase \cite{R:Gase}
and those of Simon and Agarwal \cite{R:SimonAgarwal}; in
particular, Simon and Agarwal were first to discover the closed
form expression that we rederive here, using our new point of
view.

We define the $d$-dimensional Wigner transform as
\begin{equation*}
    W_{d}(f,g)(\vec{x},\vec{\xi})=(2\pi)^{-\frac{d}{2}}\int e^{i\vec{p}\cdot\vec{\xi}}
    \overline{f\left(\frac{\vec{x}+\vec{p}}{\sqrt{2}}\right)}
    g\left(\frac{\vec{x}-\vec{p}}{\sqrt{2}}\right)\, d\vec{p}
\end{equation*}
for functions $f,g$ of $d$ variables. If $d=1$, we omit the
subscript. The Wigner transform is often restricted to the case
when $f\equiv g$, and then one writes $W_{d}(f):=W_{d}(f,f)$. More
generally, we define the extended Wigner transform, of a function
$F$ of $2d$ variables, as
\begin{equation*}
    \tilde{W}_{d}(F)(\vec{x},\vec{\xi})=(2\pi)^{-\frac{d}{2}}\int e^{i\vec{p}\cdot\vec{\xi}}
    F\left(\frac{\vec{x}+\vec{p}}{\sqrt{2}},\frac{\vec{x}-\vec{p}}{\sqrt{2}}\right)\,
    d\vec{p}.
\end{equation*}
This more general transform is fundamental for the study of LG
modes; this is our main insight and provides the central theme for
our paper.

Our main results are summarized in the following theorem.

\vspace{12pt}

\begin{theorem}
    Let $h_{j}$ denote the $j^{\text{th}}$ Hermite function, so that $h_{jk}(x,y)=h_{j}(x)h_{k}(y)$
    are the HG modes.

    \newcounter{MainResults}
    \begin{list}{(\textbf{\alph{MainResults}})}
            {\setlength{\leftmargin}{.5in}
             \setlength{\rightmargin}{.5in}
             \usecounter{MainResults}}

    \item The extended Wigner transform intertwines the creation
    and annihilation operators of the HG and LG modes.

    \item The LG modes are precisely $\tilde{W}(h_{jk})$: the extended
    Wigner transforms of the HG
    modes.

    \item The Wigner transforms of the LG modes are given by
    \begin{equation}\label{E:MainEq2}
        \begin{aligned}
        W_{2}(\tilde{W}(h_{jk})&,\tilde{W}({h_{mn}}))(\vec{x},\vec{\xi})\\
        &=\tilde{W}(h_{jm})\left(\frac{x_{1}+\xi_{2}}{\sqrt{2}},\frac{\xi_{1}-x_{2}}{\sqrt{2}}\right)
        \tilde{W}(h_{kn})\left(\frac{x_{1}-\xi_{2}}{\sqrt{2}},\frac{\xi_{1}+x_{2}}{\sqrt{2}}\right).
        \end{aligned}
    \end{equation}

    \end{list}
\end{theorem}

\vspace{12pt}

We remark that part (b) of the theorem follows from part (a) and
the fact that $\tilde{W}(h_{00})=h_{00}$.

\vspace{12pt}

In the case $(j,k)=(m,n)$, the formula (\ref{E:MainEq2}) was
proven by Simon and Agarwal, using the metaplectic representation,
and in fact their proof extends to the general case without
modification \cite{R:SimonAgarwal}. The generalization should be
useful in light of the relationship between the Wigner transform
and the Weyl quantization of observables. Indeed, for any tempered
distribution $\sigma\in\mathcal{S}^{\prime}(\Rbb^{2d})$ we may
define the Weyl quantization of $\sigma$ as the
[pseudodifferential] operator given by
\begin{equation*}
    \left(\text{Op}^{W}(\sigma)u\right)(x)
    =2^{-\frac{3d}{2}}\pi^{-d}\iint
    \exp\left(\frac{i(x-y)\xi}{\sqrt{2}}\right)\sigma\left(\frac{x+y}{\sqrt{2}},\xi\right)u(y)\,
    dy\,d\xi.
\end{equation*}
The normalization factor is chosen so that $\text{Op}^{W}(1)$ is
the identity operator. Then it is easy to check that for any
Schwartz functions $f,g\in\mathcal{S}(\Rbb^{d})$ we have
\begin{equation*}
    \langle f|\text{Op}^{W}(\sigma)g\rangle
    =2^{-d}\pi^{-\frac{d}{2}}\iint\sigma(x,\xi)W_{d}(f,g)(x,\xi)\, dx\,d\xi.
\end{equation*}
For an exposition of the Weyl calculus of pseudodifferential
operators, the reader may consult the book of Gerald Folland
\cite{R:Folland}.

\vspace{12pt}

In Section \ref{S:onedim} we review the theory of the
one-dimensional harmonic oscillator. Since this subject is very
well known, we will be brief, so the section is mostly a means of
setting up notation for the following sections. In Section
\ref{S:twodim}, we review the theory of the two-dimensional
harmonic oscillator. Our primary reference for Sections
\ref{S:onedim} and \ref{S:twodim} is the classic book of Messiah
\cite{R:Messiah}.

In Section \ref{S:twodimQQQQ} we write the $\{Q\}$-representation
in terms of complex variables, which provides a clean expression
for the LG modes in terms of Laguerre polynomials. The connection
with LG modes of paraxial light is briefly exhibited in Section
\ref{S:LGmodesparax}.

The LG modes are expressed as Wigner transforms in Section
\ref{S:Wigner}, and in Section \ref{S:WignerOfWigner} we calculate
their own Wigner transforms.

\vspace{24pt}

\section{The One-dimensional Harmonic Oscillator}\label{S:onedim}

We begin by setting up the notation for the one-dimensional
harmonic oscillator, given by
$$H=\frac{1}{2}(P^{2}+Q^{2}), \quad \text{where }[Q,P]=i.$$ We set
$$a=\frac{1}{\sqrt{2}}(Q+iP)\quad\text{and}\quad
a^{\dag}=\frac{1}{\sqrt{2}}(Q-iP)$$ so that
$$[a,a^{\dag}]=1\quad\text{and}\quad
H=\frac{1}{2}(aa^{\dag}+a^{\dag}a).$$ Moreover, we define
$$N=a^{\dag}a$$ so that $$H=N+\frac{1}{2}.$$

It is well known that $\text{Spec}(N)=\{0,1,2,\ldots\}$ consists
of nondegenerate eigenvalues and that $a^{\dag}$ and $a$ are the
``ladder operators'': the creation and annihilation operators,
respectively. The associated eigenvectors form a complete set, so
we can normalize to get an orthonormal basis of eigenvectors for
the observable $N$:
$$|0\rangle,\quad |1\rangle,\quad |2\rangle,\ldots$$
corresponding to the eigenvalues
$$0,\quad 1,\quad 2,\ldots$$

In the $\{Q\}$-Representation (that is, the Schr\"{o}dinger
representation where states are considered as functions of
position) the ladder operators are written as
$$a=\frac{1}{\sqrt{2}}\left(x+\frac{d}{dx}\right)$$ and
$$a^{\dag}=\frac{1}{\sqrt{2}}\left(x-\frac{d}{dx}\right).$$ The ground state
$h_{0}$ satisfies $$\left[\frac{d}{dx}+x\right]h_{0}(x)=0,$$ so
that
$$h_{0}(x)=\pi^{-\frac{1}{4}}e^{-\frac{1}{2}x^{2}}.$$
Using the ladder operators, we get the rest of the eigenvectors,
which are precisely the Hermite functions:
\begin{equation*}
    \begin{aligned}
    h_{n}(x)
    &=\pi^{-\frac{1}{4}}(n!)^{-\frac{1}{2}}2^{-\frac{n}{2}}\left(x-\frac{d}{dx}\right)^{n}e^{-\frac{1}{2}x^{2}}\\
    &=\pi^{-\frac{1}{4}}(n!)^{-\frac{1}{2}}2^{-\frac{n}{2}}(-1)^{n}e^{\frac{1}{2}x^{2}}\frac{d^{n}}{dx^{n}}e^{-x^{2}}\\
    &=\pi^{-\frac{1}{4}}(n!)^{-\frac{1}{2}}2^{-\frac{n}{2}}e^{-\frac{1}{2}x^{2}}H_{n}(x)
    \end{aligned}
\end{equation*}
where the $H_{n}$ are the Hermite polynomials
$$H_{n}(x)=(-1)^{n}e^{x^{2}}\frac{d^{n}}{dx^{n}}e^{-x^{2}}.$$

\vspace{24pt}

\section{The Two-Dimensional Isotropic Harmonic
Oscillator}\label{S:twodim}

In two dimensions, we have
$$H=\frac{1}{2}(P_{1}^{2}+P_{2}^{2}+Q_{1}^{2}+Q_{2}^{2})$$ with
creation and annihilation operators inherited from the
one-dimensional case:
\begin{equation*}
    \begin{aligned}
    a_{j}^{\dag}&=\frac{1}{\sqrt{2}}(Q_{j}-iP_{j}),\\
    a_{j}&=\frac{1}{\sqrt{2}}(Q_{j}+iP_{j}),\qquad j=1,2.
    \end{aligned}
\end{equation*}
These are interpreted as creation and annihilation operators,
respectively, of quanta of type $j\in\{1,2\}$. They satisfy
$$[a_{i},a_{j}]=[a_{i}^{\dag},a_{j}^{\dag}]=0,$$
$$[a_{i},a_{j}^{\dag}]=\delta_{ij}.$$

And we have the corresponding ``number operators''
$$N_{j}=a_{j}^{\dag}a_{j},\qquad j=1,2,$$ with $$N=N_{1}+N_{2}$$
representing the total number.

Now the eigenvalues for $$H=N+1$$ are $$\{1,2,3,\ldots\}$$ and the
eigenvalue $j$ has degeneracy $j$. Hence there are many possible
choices of eigenbasis. One often simply takes the basis consisting
of tensor products of the one-dimensional eigenvectors,
corresponding to the complete set of commuting observables
$\{N_{1},N_{2}\}$. The elements of this basis are given by
\begin{equation*}
    |n_{1}n_{2}\rangle=(n_{1}!n_{2}!)^{-\frac{1}{2}}a_{1}^{\dag
    n_{1}}a_{2}^{\dag n_{2}}|00\rangle.
\end{equation*}
These are precisely the HG modes; in the Schr\"{o}dinger
representation, they are tensor products of Hermite functions.

But here we will construct another basis.

\vspace{12pt}

The angular momentum operator $L$ is defined by
$$L=Q_{1}P_{2}-Q_{2}P_{1}=i(a_{1}a_{2}^{\dag}-a_{1}^{\dag}a_{2}),$$
and one may check that it is a constant of motion. We will show
that $\{N,L\}$ is another complete set of commuting observables.

Let $$A_{\pm}=\frac{1}{\sqrt{2}}(a_{1}\mp ia_{2})$$ and
$$A_{\pm}^{\dag}=\frac{1}{\sqrt{2}}(a_{1}^{\dag}\pm
ia_{2}^{\dag}).$$ Then
$$[A_{r},A_{s}]=[A_{r}^{\dag},A_{s}^{\dag}]=0\qquad\text{and}$$
$$[A_{r},A_{s}^{\dag}]=\delta_{rs} \qquad \text{for }r,s\in\{+,-\}.$$

We think of $A_{r}^{\dag}$, $A_{r}$ as creation and annihilation
operators, respectively, of quanta of type $r\in\{+,-\}$, so then
$$N_{r}=A_{r}^{\dag}A_{r}$$ represents the number of ``$r$
quanta''.

Hence the problem of finding eigenvectors common to $N_{+}$ and
$N_{-}$ is formally equivalent to finding eigenvectors common to
$N_{1}$ and $N_{2}$. So by the usual arguments we see that
$$\text{Spec}(N_{+})=\text{Spec}(N_{-})=\{0,1,2,\ldots\}$$ and
that these two observables, $N_{+}$ and $N_{-}$, form a complete
set of commuting observables: to each $(n_{+},n_{-})$ there is a
common eigenvector, denoted by $|n_{+}n_{-}\rangle$, that is
unique to within a constant.

In fact, $$A_{+}|00\rangle=A_{-}|00\rangle=0$$ and the states
$$|n_{+}n_{-}\rangle =(n_{+}!n_{-}!)^{-\frac{1}{2}}A_{+}^{\dag n_{+}}A_{-}^{\dag
n_{-}}|00\rangle$$ form a complete orthonormal eigenbasis common
to $N_{+}$ and $N_{-}$:
$$N_{+}|n_{+}n_{-}\rangle =n_{+}|n_{+}n_{-}\rangle$$ and
$$N_{-}|n_{+}n_{-}\rangle =n_{-}|n_{+}n_{-}\rangle.$$

We find that $$N=N_{+}+N_{-}$$ and that $$L=N_{+}-N_{-}.$$ Hence
$N$ and $L$ form a complete set of commuting observables.
Moreover,
$$[L,A_{\pm}^{\dag}]=\pm A_{\pm}^{\dag}$$ and
$$[L,A_{\pm}]=\mp A_{\pm},$$ so when they act upon an eigenvector
of $L$, $A_{+}^{\dag}$ and $A_{-}$ increase $L$ by one unit, and
$A_{-}^{\dag}$ and $A_{+}$ decrease $L$ by one unit. So it is
natural to consider $N_{+}$ as the number of particles with
positive charge, $N_{-}$ as the number of particles with negative
charge, and $L$ as the total charge (to within a constant).

\vspace{24pt}

\section{The Two-Dimensional Isotropic Harmonic Oscillator in the
$\{Q\}$-Representation}\label{S:twodimQQQQ}

In the $\{Q\}$-representation, we write
\begin{equation}\label{E:littlea}
    \begin{aligned}
    a_{1}&=\frac{1}{\sqrt{2}}\left(x+\frac{\partial}{\partial x}\right),\qquad
    a_{2}=\frac{1}{\sqrt{2}}\left(y+\frac{\partial}{\partial y}\right),\\
    a_{1}^{\dag}&=\frac{1}{\sqrt{2}}\left(x-\frac{\partial}{\partial
    x}\right),\qquad
    a_{2}^{\dag}=\frac{1}{\sqrt{2}}\left(y-\frac{\partial}{\partial
    y}\right),
    \end{aligned}
\end{equation}
so that we have
\begin{equation}\label{E:BigA}
    \begin{aligned}
    A_{+}&=\frac{1}{2}\left(x+\frac{\partial}{\partial x}-i\left(y+\frac{\partial}{\partial
    y}\right)\right),\\
    A_{-}&=\frac{1}{2}\left(x+\frac{\partial}{\partial x}+i\left(y+\frac{\partial}{\partial
    y}\right)\right),\\
    A_{+}^{\dag}&=\frac{1}{2}\left(x-\frac{\partial}{\partial x}+i\left(y-\frac{\partial}{\partial
    y}\right)\right),\\
    A_{-}^{\dag}&=\frac{1}{2}\left(x-\frac{\partial}{\partial x}-i\left(y-\frac{\partial}{\partial y}\right)\right).
    \end{aligned}
\end{equation}
In complex notation, these take a simple form. We write $z=x+iy$,
so that
\begin{equation*}
    \frac{\partial}{\partial
    z}=\frac{1}{2}\left(\frac{\partial}{\partial
    x}+\frac{1}{i}\frac{\partial}{\partial y}\right)\qquad
    \text{and}\qquad \frac{\partial}{\partial
    \overline{z}}=\frac{1}{2}\left(\frac{\partial}{\partial
    x}-\frac{1}{i}\frac{\partial}{\partial y}\right).
\end{equation*}

Then

\begin{equation*}
    \begin{aligned}
    A_{+}&=\frac{1}{2}\overline{z}+\frac{\partial}{\partial
    z},\qquad A_{-}=\frac{1}{2}z+\frac{\partial}{\partial
    \overline{z}},\\
    A_{+}^{\dag}&=\frac{1}{2}z-\frac{\partial}{\partial
    \overline{z}},\qquad A_{-}^{\dag}=\frac{1}{2}\overline{z}-\frac{\partial}{\partial z}.
    \end{aligned}
\end{equation*}
Moreover, we can write
\begin{equation*}
    A_{+}^{\dag}=-e^{\frac{1}{2}z\overline{z}}\frac{\partial}{\partial\overline{z}}e^{-\frac{1}{2}z\overline{z}}
    \qquad\text{and}\qquad
    A_{-}^{\dag}=-e^{\frac{1}{2}z\overline{z}}\frac{\partial}{\partial
    z}e^{-\frac{1}{2}z\overline{z}}.
\end{equation*}

The ground state is given by
$$u_{0}(z,\overline{z})=\pi^{-\frac{1}{2}}e^{-\frac{1}{2}z\overline{z}}\equiv\langle z,\overline{z}|00\rangle,$$
and of course we get all other eigenvectors by applying the
creation operators to this. These are precisely the LG modes.
Explicitly,
\begin{equation*}
    \begin{aligned}
    \langle z,\overline{z}|\npl\nmi\rangle
    &=\langle z,\overline{z}|(\npl!\nmi!)^{-\frac{1}{2}}A_{+}^{\dag\npl}A_{-}^{\dag\nmi}|00\rangle\\
    &=\pi^{-\frac{1}{2}}(\npl!\nmi!)^{-\frac{1}{2}}(-1)^{\npl+\nmi}
    e^{\frac{1}{2}z\zbar}\left(\frac{\partial}{\partial\zbar}\right)^{\npl}
    \left(\frac{\partial}{\partial z}\right)^{\nmi}e^{-z\zbar}.
    \end{aligned}
\end{equation*}
We will now formulate this in terms of Laguerre polynomials.

First suppose that $\npl\geq\nmi$. Then
\begin{equation*}
    \begin{aligned}
    \langle z,\overline{z}|\npl\nmi\rangle
    &=\pi^{-\frac{1}{2}}(\npl!\nmi!)^{-\frac{1}{2}}(-1)^{\npl+\nmi}
    e^{\frac{1}{2}z\zbar}\left(\frac{\partial}{\partial
    z}\right)^{\nmi}\left[(-z)^{\npl}e^{-z\zbar}\right]\\
    &=\pi^{-\frac{1}{2}}(\npl!\nmi!)^{-\frac{1}{2}}(-1)^{\nmi}
    e^{\frac{1}{2}z\zbar}\zbar^{(\nmi-\npl)}\left(\frac{\partial}{\partial
    (z\zbar)}\right)^{\nmi}\left[(z\zbar)^{\npl}e^{-z\zbar}\right].
    \end{aligned}
\end{equation*}
We define the Laguerre polynomials by
$$L^{\alpha}_{n}(x)=\frac{x^{-\alpha}e^{x}}{n!}\frac{d^{n}}{dx^{n}}\left(e^{-x}x^{n+\alpha}\right)$$
for $n\geq 0$ and $\alpha>-1$. Then we see that, when
$\npl\geq\nmi$,
\begin{equation*}
    \langle z,\overline{z}|\npl\nmi\rangle=\pi^{-\frac{1}{2}}\left(\frac{\nmi!}{\npl!}\right)^{\frac{1}{2}}
    (-1)^{\nmi}z^{(\npl-\nmi)}e^{-\frac{1}{2}z\zbar}L^{\npl-\nmi}_{\nmi}(z\zbar).
\end{equation*}

On the other hand, if $\npl\leq\nmi$, similar arguments show that
\begin{equation*}
    \langle z,\overline{z}|\npl\nmi\rangle=\pi^{-\frac{1}{2}}\left(\frac{\npl!}{\nmi!}\right)^{\frac{1}{2}}
    (-1)^{\npl}\zbar^{(\nmi-\npl)}e^{-\frac{1}{2}z\zbar}L^{\nmi-\npl}_{\npl}(z\zbar).
\end{equation*}
We summarize this in the following theorem:

\vspace{12pt}

\begin{theorem}\label{T:laguerrestates}
    We consider the states $|\npl\nmi\rangle$
    in the $\{Q\}$-representation with position coordinates $(x,y)\in\Rbb^{2}$, and we write
    $z=x+iy$. Then
    \begin{equation*}
        \langle z,\overline{z}|\npl\nmi\rangle=
    \begin{cases}
        \pi^{-\frac{1}{2}}\left(\frac{\nmi!}{\npl!}\right)^{\frac{1}{2}}
        (-1)^{\nmi}z^{(\npl-\nmi)}e^{-\frac{1}{2}z\zbar}L^{\npl-\nmi}_{\nmi}(z\zbar)
        &\text{if }\npl\geq\nmi,\text{ and}\\
        \pi^{-\frac{1}{2}}\left(\frac{\npl!}{\nmi!}\right)^{\frac{1}{2}}
        (-1)^{\npl}\zbar^{(\nmi-\npl)}e^{-\frac{1}{2}z\zbar}L^{\nmi-\npl}_{\npl}(z\zbar)&\text{if }\npl\leq\nmi.
        \end{cases}
    \end{equation*}
\end{theorem}

\vspace{24pt}

\section{Comparison With LG Modes of Paraxial
Light}\label{S:LGmodesparax}

The time-harmonic field amplitudes of LG optical modes are
expressed in cylindrical coordinates, with propagation coordinate
$z$ along the beam axis, as
\begin{equation*}
    \tilde{u}_{p,\ell}(\mathbf{r})\propto
    e^{-i\Phi(\mathbf{r})}e^{-\frac{r^{2}}{w(z)^{2}}}\left(\frac{r\sqrt{2}}{w(z)}\right)^{|\ell|}
    L^{|\ell|}_{p}\left(\frac{2r^{2}}{w(z)^{2}}\right)
\end{equation*}
where the phase is
\begin{equation*}
    \Phi(\mathbf{r})=\ell\phi-kz+\frac{kr^{2}}{2R(z)}-(2p+\ell+1)\Psi(z)
\end{equation*}
and where again $L^{|\ell|}_{p}$ is the Laguerre polynomial with
radial and azimuthal indices $p$ and $\ell$, respectively. The
term $\Psi(z)=\tan^{-1}(\frac{z}{z_{R}})$ is part of the Gouy
phase, $R(z)=\frac{z_{R}^{2}+z^{2}}{z}$ is the radius of
curvature, $w(z)=w_{0}\sqrt{1+(\frac{z}{z_{R}})^{2}}$ is the beam
waist, and $z_{R}=\frac{1}{2}kw_{0}^{2}$ is the Rayleigh range. We
see that the total phase evolves helically along the propagation
coordinate $z$. For more on this, see, for example,
\cite{R:OpticalAM}, \cite{R:PAC07}, \cite{R:Siegman}.

In the transversal plane at the beam waist (z=0), we get
\begin{equation*}
    \tilde{u}_{p,\ell}(\mathbf{r})\propto
    e^{-i\ell\phi}e^{-\frac{r^{2}}{w_{0}^{2}}}\left(\frac{r\sqrt{2}}{w_{0}}\right)^{|\ell|}
    L^{|\ell|}_{p}\left(\frac{2r^{2}}{w_{0}^{2}}\right),
\end{equation*}
which is essentially what we have in Theorem
\ref{T:laguerrestates}.

\vspace{24pt}

\section{LG Modes Are Wigner Transforms}\label{S:Wigner}

Physicists in the optics community recently computed the Wigner
transforms of LG modes \cite{R:Gase}, \cite{R:SimonAgarwal}.
Surprisingly, the states $|\npl\nmi\rangle$ are already themselves
Wigner transforms of Hermite functions. The Hermite functions, as
in Section \ref{S:onedim}, are given by
\begin{equation*}
    h_{j}(x)=\pi^{-\frac{1}{4}}(j!)^{-\frac{1}{2}}2^{-\frac{j}{2}}(-1)^{j}e^{\frac{1}{2}x^{2}}
    \frac{d^{j}}{dx^{j}}e^{-x^{2}},
\end{equation*}
which, as we know, satisfy the orthogonality relations
\begin{equation*}
    \langle
    h_{j}|h_{k}\rangle=\delta_{jk}.
\end{equation*}
And we continue to write the HG modes as
$$h_{jk}(x,y)=h_{j}(x)h_{k}(y).$$

Now, for reference, we list the standard properties of the Wigner
transform:

\begin{equation}
    W(f,g)=\overline{W(g,f)},
\end{equation}
\begin{equation}
    \int W(f,g)(x,\xi)\, d\xi=\sqrt{2\pi}\overline{f\left(\frac{x}{\sqrt{2}}\right)}g\left(\frac{x}{\sqrt{2}}\right),
\end{equation}
\begin{equation}
    \iint W(f,g)(x,\xi)\,
    dx\,d\xi=2\sqrt{\pi}\langle f|g\rangle,
\end{equation}
\begin{equation}
    \int W(f,g)(x,\xi)\, dx=\sqrt{2\pi}\overline{\hat{f}\left(\frac{\xi}{\sqrt{2}}\right)}
    \hat{g}\left(\frac{\xi}{\sqrt{2}}\right),
\end{equation}
where $\hat{f}(\xi)=\frac{1}{\sqrt{2\pi}}\int e^{-ix\xi}f(x)\, dx$
is the Fourier transform, and (``Moyal's identity'')
\begin{equation}
    \langle W(f_{1},g_{1})| W(f_{2},g_{2})\rangle
    =\overline{\langle f_{1}|f_{2}\rangle}\langle g_{1}|g_{2}\rangle.
\end{equation}

\vspace{12pt}

These properties are well known and are easily checked. But there
is another elementary property which seems to be a new
observation:

\vspace{12pt}

\begin{theorem}\label{T:intertwines}
    The extended Wigner transform intertwines the creation
    operators of the HG and LG modes, and it intertwines the
    annihilation operators of the HG and LG modes:
    \begin{align*}
        A_{+}^{\dag}\tilde{W} &= \tilde{W}a_{1}^{\dag},\qquad\qquad
        A_{-}^{\dag}\tilde{W} = \tilde{W}a_{2}^{\dag},\\
        A_{+}\tilde{W} &= \tilde{W}a_{1},\qquad\qquad
        A_{-}\tilde{W} = \tilde{W}a_{2}.
    \end{align*}
\end{theorem}

\vspace{12pt}

The proof is a direct calculation, using the expressions
(\ref{E:littlea}) and (\ref{E:BigA}). As a first application, it
is now easy to see that the extended Wigner transform commutes
with the operators $H$ and $N$.

\vspace{12pt}

Additionally using the fact that $\tilde{W}(h_{00})=h_{00}$, we
have the following corollary:

\vspace{12pt}

\begin{corollary}\label{C:stateWig}
    We consider the LG modes $|\npl\nmi\rangle$
    in the $\{Q\}$-representation with position coordinates $(x,y)\in\Rbb^{2}$. Then we have
    $$\langle x,y|\npl\nmi\rangle=W(h_{\npl},h_{\nmi})(x,y).$$
\end{corollary}

\vspace{12pt}

Combining this with Theorem \ref{T:laguerrestates}, we have a
formula:

\vspace{12pt}

\begin{corollary}\label{C:FollandThm}
    Suppose $x,y\in\Rbb$, and let $z=x+iy$. Then
    \begin{equation*}
        W(h_{j},h_{k})(x,y)=
        \begin{cases}
        \pi^{-\frac{1}{2}}\left(\frac{k!}{j!}\right)^{\frac{1}{2}}
        (-1)^{k}z^{j-k}e^{-\frac{1}{2}z\zbar}L^{j-k}_{k}(z\zbar)
        &\text{if }j\geq k, \text{ and}\\
        \pi^{-\frac{1}{2}}\left(\frac{j!}{k!}\right)^{\frac{1}{2}}
        (-1)^{j}\zbar^{k-j}e^{-\frac{1}{2}z\zbar}L^{k-j}_{j}(z\zbar)
        &\text{if }j\leq k.
        \end{cases}
    \end{equation*}
\end{corollary}

\vspace{12pt}

Corollary \ref{C:FollandThm} by itself is not new. For example,
two independent proofs may be found in Folland's book
\cite{R:Folland}, although we caution the reader that his
statement of the result contains a misprint.

\vspace{12pt}

So we have yet another way of understanding the orthogonality
relations of the states $|n_{+}n_{-}\rangle$: as being inherited
from the orthogonality of the Hermite functions. Explicitly,
\begin{equation*}
    \begin{aligned}
    \langle m_{+}m_{-}|n_{+}n_{-}\rangle
    &=\langle
    W(h_{m_{+}},h_{m_{-}})|W(h_{n_{+}},h_{n_{-}})\rangle\\
    &=\langle h_{m_{+}}|h_{n_{+}}\rangle\langle h_{m_{-}}|h_{n_{-}}\rangle\\
    &=\delta_{m_{+}n_{+}}\delta_{m_{-}n_{-}}.
    \end{aligned}
\end{equation*}

\vspace{12pt}

As another application of Corollary \ref{C:stateWig}, we can think
of the LG modes $|\npl\nmi\rangle$ as being ``interference terms''
resulting from the quadratic nature of the Wigner transform. If we
apply the Wigner transform to sums of Hermite functions, we get
interference terms that are LG modes with nonzero angular
momentum. In fact, we have the following polarization identity:
\begin{equation*}
    \begin{aligned}
    \langle
    x,y|n_{+}n_{-}\rangle&=\frac{1}{4}W(h_{n_{+}}+h_{n_{-}})(x,y)
    -\frac{1}{4}W(h_{n_{+}}-h_{n_{-}})(x,y)\\
    &\qquad\qquad+\frac{i}{4}W(h_{n_{+}}-ih_{n_{-}})(x,y)
    -\frac{i}{4}W(h_{n_{+}}+ih_{n_{-}})(x,y).
    \end{aligned}
\end{equation*}

\vspace{12pt}

One may think of the Wigner transform as being a kind of
``product''. Just as Hermite functions give rise to
two-dimensional HG modes, via the tensor product, Hermite
functions also give rise to two-dimensional LG modes, via the
Wigner transform. To complete this series of relationships, Simon
and Agarwal \cite{R:SimonAgarwal} noted that the two-dimensional
HG and LG modes are unitarily related by the operator
$$\exp\left(\frac{i\pi}{4}\hat{T}_{1}\right),\qquad\text{where }\hat{T}_{1}=xy-\frac{\partial^{2}}{\partial x\partial y}.$$

Alternatively, we can use the Wigner transform to again show that
the HG and LG modes are unitarily related. For functions $F$ of
two variables, we consider the extended Wigner transform of $F$:
\begin{equation*}
    \tilde{W}F(x,y)=\frac{1}{\sqrt{2\pi}}\int e^{ipy}
    F\left(\frac{x+p}{\sqrt{2}},\frac{x-p}{\sqrt{2}}\right)\, dp.
\end{equation*}
If we write the $\frac{\pi}{4}$-rotation operator as
\begin{align*}
    (R^{*}_{\frac{\pi}{4}}F)(x,p)&=F(R_{\frac{\pi}{4}}(x,p))\\
    &=F\left(\frac{x-p}{\sqrt{2}},\frac{x+p}{\sqrt{2}}\right),
\end{align*}
and if we write the partial Fourier transform as
\begin{equation*}
    (\mathcal{F}_{2}F)(x,y)=\frac{1}{\sqrt{2\pi}}\int
    e^{-ipy}F(x,p)\, dp,
\end{equation*}
then we may write the extended Wigner transform as
\begin{equation}\label{E:WigUnitary}
    \tilde{W}F(x,y)=\left(\mathcal{F}_{2}R^{*}_{\frac{\pi}{4}}F\right)(x,y).
\end{equation}
This is clearly a unitary operator, so, in particular, we again
see that the HG modes $h_{jk}$ are unitarily related to the LG
modes $\tilde{W}(h_{jk})$.

\vspace{12pt}

The results of this section are surprising, despite their
mathematical simplicity. We are accustomed to thinking of the
Wigner transform of a function as being a phase space
representation of the function, but here the relevant Wigner
transform is actually a function of the position vector
$(x,y)\in\Rbb^{2}$, as it is an LG mode. So if we use the Wigner
transform to study the phase space properties of LG modes, we are
really taking Wigner transforms of Wigner transforms. Gase
\cite{R:Gase}, followed by Simon and Agarwal
\cite{R:SimonAgarwal}, already recently derived expressions for
Wigner transforms of LG modes (Gase in terms of a quadruple sum,
and Simon and Agarwal in terms of a closed form involving no
summation at all). We will do the same in the next section, but
now using our new point of view.

\vspace{24pt}

\section{Wigner Transforms of LG Modes}\label{S:WignerOfWigner}

We now proceed to compute the Wigner transforms of the LG modes
$\tilde{W}(h_{jk})$. In light of (\ref{E:WigUnitary}), the first
step is to consider the action of the Wigner transform on partial
Fourier transforms.

\vspace{12pt}

\begin{lemma}
    \begin{equation*}
    W_{2}(\mathcal{F}_{2}F,\mathcal{F}_{2}G)(\vec{x},\vec{\xi})
    =\frac{1}{2\pi}\iint e^{i\xi_{1}s-ix_{2}t}
    \overline{F\left(\frac{x_{1}+s}{\sqrt{2}},\frac{-\xi_{2}-t}{\sqrt{2}}\right)}
    G\left(\frac{x_{1}-s}{\sqrt{2}},\frac{-\xi_{2}+t}{\sqrt{2}}\right)\,
    ds\,dt.
    \end{equation*}
\end{lemma}

\vspace{12pt}

This lemma follows from a direct calculation.

\vspace{12pt}

We now apply the lemma to
\begin{equation*}
    F=R^{*}_{\frac{\pi}{4}}h_{jk}\qquad\text{and}\qquad
    G=R^{*}_{\frac{\pi}{4}}h_{mn}.
\end{equation*}
That is,
\begin{equation*}
    F(\vec{x})=h_{j}\left(\frac{x_{1}-x_{2}}{\sqrt{2}}\right)h_{k}\left(\frac{x_{1}+x_{2}}{\sqrt{2}}\right)
\end{equation*}
and similarly for $G$. Hence the Wigner transforms of the LG modes
are given by
\begin{align*}
    W_{2}(\tilde{W}(h_{jk}),\tilde{W}(h_{mn}))(\vec{x},\vec{\xi})\\
    =\frac{1}{2\pi}\iint e^{i\xi_{1}s-ix_{2}t}
    &h_{j}\left(\frac{(x_{1}+s)+(\xi_{2}+t)}{2}\right)
    h_{k}\left(\frac{(x_{1}+s)-(\xi_{2}+t)}{2}\right)\\
    &h_{m}\left(\frac{(x_{1}-s)+(\xi_{2}-t)}{2}\right)
    h_{n}\left(\frac{(x_{1}-s)-(\xi_{2}-t)}{2}\right)\, ds\,dt.
\end{align*}
We may simplify this integral by again rotating the variables by
$\frac{\pi}{4}$. We state the result as a theorem.

\vspace{12pt}

\begin{theorem}\label{T:WignerofWigner}
The Wigner transforms of the LG modes are products of LG modes,
given by the formula
    \begin{align*}
        W_{2}(\tilde{W}(h_{jk}),&\tilde{W}(h_{mn}))(\vec{x},\vec{\xi})\\
        &=\tilde{W}(h_{jm})\left(\frac{x_{1}+\xi_{2}}{\sqrt{2}},\frac{\xi_{1}-x_{2}}{\sqrt{2}}\right)
        \tilde{W}(h_{kn})\left(\frac{x_{1}-\xi_{2}}{\sqrt{2}},\frac{\xi_{1}+x_{2}}{\sqrt{2}}\right).
    \end{align*}
\end{theorem}

\vspace{12pt}

In the special case $(j,k)=(m,n)$, we have the following formula
of Simon and Agarwal \cite{R:SimonAgarwal}. It is now proven by
combining Corollary \ref{C:FollandThm} with Theorem
\ref{T:WignerofWigner}.

\vspace{12pt}

\begin{corollary}
For
$(\vec{x},\vec{\xi})=(x_{1},x_{2},\xi_{1},\xi_{2})\in\Rbb^{2}\times\Rbb^{2}$,
we let
$$Q_{0}=\frac{1}{2}(|\vec{x}|^{2}+|\vec{\xi}|^{2})\qquad\text{and}\qquad Q_{2}=(x_{1}\xi_{2}-x_{2}\xi_{1}).$$
Then we have
    \begin{equation*}
        W_{2}(\tilde{W}(h_{jk}),\tilde{W}(h_{jk}))(\vec{x},\vec{\xi})
        =\pi^{-1}(-1)^{j+k}e^{-Q_{0}}L^{0}_{j}(Q_{0}+Q_{2})L^{0}_{k}(Q_{0}-Q_{2}).
    \end{equation*}
\end{corollary}

\vspace{12pt}

For comparison, we also give the Wigner transforms of HG modes:
\begin{equation*}
    W_{2}(h_{jk},h_{mn})(\vec{x},\vec{\xi})
    =\tilde{W}(h_{jm})(x_{1},\xi_{1})\tilde{W}(h_{kn})(x_{2},\xi_{2}).
\end{equation*}
In the particular case $(j,k)=(m,n)$, if we let
$$Q_{3}=\frac{1}{2}(x_{1}^{2}-x_{2}^{2}+\xi_{1}^{2}-\xi_{2}^{2}),$$
then we have the formula
\begin{equation*}
    W_{2}(h_{jk},h_{jk})(\vec{x},\vec{\xi})
    =\pi^{-1}(-1)^{j+k}e^{-Q_{0}}L^{0}_{j}(Q_{0}+Q_{3})L^{0}_{k}(Q_{0}-Q_{3}).
\end{equation*}

\vspace{12pt}

\textbf{Acknowledgements:} The author's interest in this topic is
due to a fascinating talk given by J. B. Rosenzweig, entitled
``Accelerator Physics: New Light Sources and Medical Imaging
Techniques''. The author also wishes to thank M. Hitrik and R.
Simon for their helpful comments.

\end{document}